\documentclass[sigconf]{acmart}

\usepackage{booktabs} 
\usepackage[ruled]{algorithm2e} 

\SetAlFnt{\small}
\SetAlCapFnt{\small}
\SetAlCapNameFnt{\small}
\SetAlCapHSkip{0pt}
\usepackage{epsfig}
\usepackage{epstopdf}
\usepackage{multirow}
\usepackage{balance}

\newlength\mylen
\newcommand\myinput[1]{%
  \settowidth\mylen{\KwIn{}}%
  \setlength\hangindent{\mylen}%
  \hspace*{\mylen}#1\\}

\copyrightyear{2018} 
\acmYear{2018} 
\setcopyright{acmcopyright}
\acmConference[ICPP 2018]{47th International Conference on Parallel Processing}{August 13--16, 2018}{Eugene, OR, USA}
\acmBooktitle{ICPP 2018: 47th International Conference on Parallel Processing, August 13--16, 2018, Eugene, OR, USA}
\acmPrice{15.00}
\acmDOI{10.1145/3225058.3225101}
\acmISBN{978-1-4503-6510-9/18/08}

\begin{document}

\title[Scalable Single Pulse Identification and Classification]{Scalable Solutions for Automated Single Pulse Identification and Classification in Radio Astronomy}

\author{Thomas R Devine}
\affiliation{
	\department{Center for Gravitational Waves and Cosmology}
	\institution{West Virginia University}
	\city{Morgantown}
	\state{West Virginia}
	}
\additionalaffiliation{
	\institution{Fairmont State University}
	\city{Fairmont}
	\state{West Virginia}
	}
\email{tdevine4@mix.wvu.edu}

\author{Katerina Goseva-Popstojanova}
\affiliation{
	\department{Center for Gravitational Waves and Cosmology}
	\institution{West Virginia University}
	\city{Morgantown}
	\state{West Virginia}
	}
\email{katerina.goseva@mail.wvu.edu}

\author{Di Pang}
\affiliation{
	\department{Center for Gravitational Waves and Cosmology}
	\institution{West Virginia University}
	\city{Morgantown}
	\state{West Virginia}
	}
\email{dipang@mix.wvu.edu}

\begin{abstract}

Data collection for scientific applications is increasing exponentially and is forecasted to soon reach peta- and exabyte scales. Applications which process and analyze scientific data must be scalable and focus on execution performance to keep pace. In the field of radio astronomy, in addition to increasingly large datasets, tasks such as the identification of transient radio signals from extrasolar sources are computationally expensive. We present a scalable approach to radio pulsar detection written in Scala that parallelizes candidate identification to take advantage of in-memory task processing using Apache Spark on a YARN distributed system. Furthermore, we introduce a novel automated multiclass supervised machine learning technique that we combine with feature selection to reduce the time required for candidate classification. Experimental testing on a Beowulf cluster with 15 data nodes shows that the parallel implementation of the identification algorithm offers a speedup of up to 5X that of a similar multithreaded implementation. Further, we show that the combination of automated multiclass classification and feature selection speeds up the execution performance of the RandomForest machine learning algorithm by an average of 54\% with less than a 2\% average reduction in the algorithm's ability to correctly classify pulsars. The generalizability of these results is demonstrated by using two real-world radio astronomy data sets.

\end{abstract}

\thispagestyle{plain}



\maketitle

\section{Introduction}
With the advancement of data collection technologies, the commercial, government, and scientific sectors are flooded with more data than a single microprocessor could possibly process. In radio astronomy, as telescopes become larger and more sensitive, data collection rates are approaching peta- and exascale ranges \cite{bigastro}. The proposed Square Kilometer Array, which will combine 2,000 dishes and a million antennae over a collection area covering one million square meters, will be capable of collecting 160 terabytes of radio data a day \cite{6409014}. For effective classification of very large data sets, algorithms must be time-efficient and scalable.

This paper builds on our previous work, which focused on identifying and classifying transient radio signals received by large radio telescopes  \cite{mnras}. Transient radio signals are short bursts of radiation detected at radio frequencies from sources such as pulsars and rotating radio transients (RRATs). Pulsars are extremely dense, rapidly spinning stars which emit radiation from their magnetic poles \cite{handbook}. If those emission beams sweep past the Earth, they can be detected as ``pulses'', similar to viewing the bright pulses of light from a lighthouse at night. RRATs are special pulsars that emit sporadically \cite{2006Natur.439..817M}.

As our radio astronomy data sets grew in size and the granularity of our focus became finer, we found several key bottlenecks in the process of transient signal identification and classification. In this paper, we present the parallelized solution we developed to overcome some of these bottlenecks with empirical evidence of its successful application to two large real-world data sets from radio sky surveys. The research questions which guided our experimental research are divided into three categories:
\begin{enumerate}
\item the \textit{scalability} of signal identification (RQ 1 -- 2),
\item the performance impact of \textit{multiclass classification} (RQ 3 -- 5), and 
\item the performance impact of \textit{feature subset selection} (RQ 6 -- 7).
\end{enumerate}

We measure success by the ability to classify instances both correctly and efficiently. Throughout this paper, we use the term \textit{classification performance} to describe a classifier's ability to classify instances correctly and \textit{execution performance} to describe efficiency with respect to timeliness. With the looming prospect of peta- and even exa-scale data collection in radio astronomy (and many other fields), the execution performance of algorithms which process or mine data becomes increasingly important.

To achieve scalability and improve execution performance, we propose D-RAPID, a software solution for searching very large radio data sets by leveraging automated data analytics and parallel data processing on distributed systems. D-RAPID offers three main contributions. First, we scale-up a modified version of the identification algorithm we first presented in \cite{mnras} to run in parallel using Apache Spark on a Hadoop YARN distributed system and show that it outperforms its multithreaded counterpart by processing data up to five times faster. Second, we offer a novel automated multiclass pulsar classification technique, which improves the execution performance of RandomForest, an ensemble tree machine learning algorithm, by 47\% with less than a 2\% reduction in classification performance, and also more accurately classifies cases missed by binary classification.

\section{Related Work}

Many recent works have applied data analytics to classification problems from radio astronomy (see \cite{mnras} and \cite{di} and references). While execution performance for scientific applications with Big Data has been considered in other fields (see \cite{Aji:2013:EGM:2493123.2462915}, \cite{Allen:2003:EAG:2747773.2747780}, \cite{D'Agostino:2014:DPF:2683593.2683652}, \cite{Qiu2016}) and other areas of astronomy (see  \cite{bue2014astronomical}, \cite{6898782}, \cite{6898791}, \cite{6898733}), most of the literature from pulsar astronomy, including our previous works\cite{mnras} and \cite{di}, are focused on classification performance with little consideration for execution performance.

It appears that \cite{Zhang:2015:SCM:2877966.2878153}, which used Apache Spark to develop a toolkit called Kira SE to parallelize source extraction for astronomy image processing, is the only work to account for execution performance in radio astronomy applications. In \cite{Zhang:2015:SCM:2877966.2878153}, several experiments processing a 1 TB dataset show that the distributed version of Kira SE outperformed a C coded version and a distributed Apache Spark platform on the Amazon cloud is a competitive alternative to using supercomputers. Our approach differs from \cite{Zhang:2015:SCM:2877966.2878153} by being more ``white box'', i.e., we address specific performance bottlenecks in the process of candidate identification and classification.

\section{Background}
\label{sect:background}

In radio astronomy, a typical pulsar search proceeds in four phases: signal collection, dedispersion, single pulse or periodicity searching, and candidate inspection. First, radio telescopes receive signals as a time-series of voltages. Second, the signals are corrected for frequency-dependent time delays introduced by traveling through the interstellar medium. The correction process, known as dedispersion, takes into account the dispersion measure (DM), or integrated number of free electrons between the source and receiver \cite{handbook}. In the third phase, either periodicity or single pulse searches occur at a number of trial DM values using specialized software, such as PRESTO\cite{2001PhDT.......123R}. While periodicity searches involve transforming and ``folding'' the dedispersed data to identify signals with regular periods \cite{1996A&AS..117..197L}, single pulse searches avoid these steps to retain sensitivity to signals with weaker signal-to-noise ratios (SNRs) or sporadically emitted signals, such as those produced by RRATs. Finally, the candidate plots produced by either search are inspected manually or classified by various machine learning techniques.

For machine learning algorithms to perform candidate classification, possible candidates must first be identified and characterized by extracting features from the data. As data sets become very large, candidate identification is increasingly processor intensive and becomes an execution performance bottleneck in the process. Our previous work focused on the identification and classification of dispersed pulse groups (DPGs) \cite{mnras}, and developing clustering algorithms to identify groups of single pulse events \cite{di} in candidate plots generated by running PRESTO's single pulse processing tool, $single\_pulse\_search.py$.

\begin{figure}[h]
	\centering
	\includegraphics[width=0.9\linewidth]{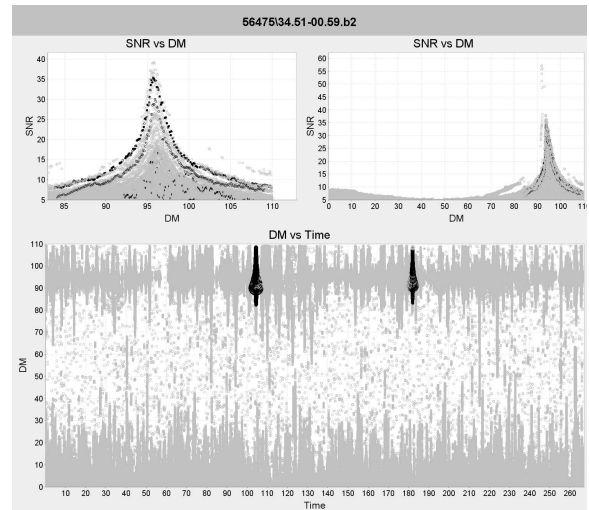}
	\caption{The known pulsar B1853+01 identified by a single pulse search.}
	\label{fig:spplot1}
\end{figure}

The primary focus of this work is to identify and classify \textit{single pulses} (SPs), which we define as clusters of single pulse events (SPEs) possibly representative of pulsar emissions. Figure~\ref{fig:spplot1} shows a customized single pulse search candidate we generated for the known pulsar, B1853+01. In the SNR vs DM (top) and DM vs time subplots (bottom) of Figure~\ref{fig:spplot1}, each individual data point represents a SPE. Single pulses are collections of SPEs appearing as clusters of points in the DM vs time space that have distinct peaks when viewed in the SNR vs DM space. Narrowing our focus from DPGs to SPEs allows for the identification and classification of signals that are much fainter, or that might be obscured by RFI, when viewed on a larger scale. In Figure~\ref{fig:spplot1}, we emphasize two individual single pulses, shown in black. In all three subplots, only the SPEs contained by the emphasized single pulses are colored black, all others are greyed out.

\section{Application Data}
\label{sect:spdata}

The data used for the experiments in this work came from two sky surveys, GBT350Drift and PALFA. GBT350Drift refers to a 350-MHz drift-scan survey performed with the Green Bank Telescope (GBT) from May through August in 2011. While the GBT was immobilised for refurbishing, the receivers remained active and collected data at 350 MHz as the sky passed through the beam of the telescope \cite{gbt350drift}. PALFA is a long-term pulsar survey of the Galactic plane using the Arecibo L-band Feed Array \cite{0004-637X-637-1-446} with a seven-beam receiver operating at 1.4 GHz with 0.3 GHz bandwidth.

Supervised machine learning depends on knowing the class values of all of the training data \textit{a priori}, which required the creation of fully labeled benchmark data sets. We derived the benchmark dataset for the GBT350Drift from the benchmark created in \cite{mnras}. From the 317 separate observations of 48 distinct pulsars in that data set, we were able to identify 5,204 single pulses. We combined these positive examples with 100,000 confirmed negative examples to create a single pulse benchmark for the GBT350Drift data. 

To identify and label single pulses from known pulsars in the PALFA data set, we used the ATNF Pulsar Catalog\cite{psrcat-paper} and RRATalog\footnote{http://astro.phys.wvu.edu/rratalog} to search our data for single pulses in the immediate vicinity of all known pulsars and RRATs and then manually inspected them for confirmation. This resulted in 3,170 single pulses from 98 known pulsars and RRATs, which were combined with 100,000 randomly selected and manually verified negative examples of single pulses from noise or RFI.

\section{Scalable Single Pulse Identification and Classification}
\label{sect:approach}

\begin{figure*}[ht]
	\centering
	\includegraphics[width=0.6\linewidth]{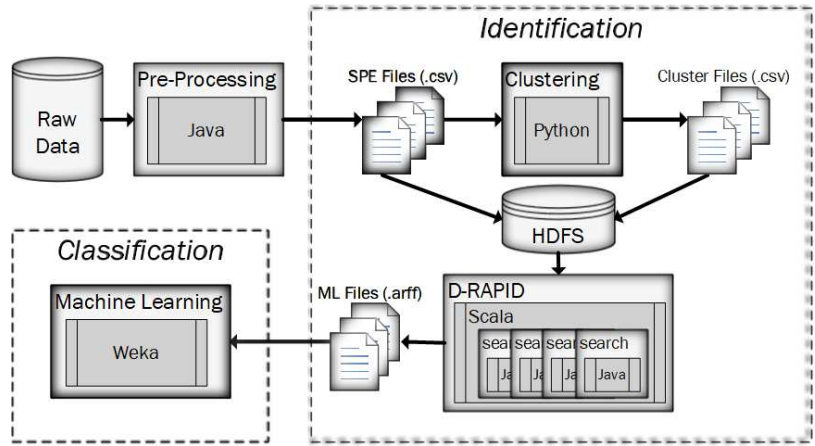}
	\caption{The scientific workflow of our scalable approach to single pulse identification and classification.}
	\label{fig:singlepulseflow}
\end{figure*}

Our approach to single pulse identification and classification consists of four stages, as depicted by our \textit{scientific workflow} in Figure~\ref{fig:singlepulseflow} \cite{Ludascher:2006:SWM:1148437.1148454}. (Note that by ``raw data'', we refer to data that has already been processed through the first three phases of a single pulse search described in Section~\ref{sect:background}.) In stage one, the raw data are pre-processed into a series of SPE files. In stage two, the SPE files are fed into a customized DBSCAN clustering algorithm \cite{di} which solves problems specific to radio astronomy clustering, such as the merging of clusters from one single pulse that appear disparate due to artifacts of data processing. After identifying clusters of associated SPEs in the DM vs Time space, we extract characteristic features from them to create a series of files describing the clusters found. The SPE and cluster files are then uploaded to the Hadoop Distributed File System (HDFS). In stage three, the Scala driver for our Distributed Recursive Algorithm for Peak IDentification (D-RAPID) distributes and executes workers to search the clusters for  single pulses, extracts features from all identified single pulses, and saves the results back to the HDFS as machine learning (ML) files. Finally, the ML files are aggregated and used for classification in stage four. This paper is focused on stages three and four with the goal of improving the execution performance of both identification (\ref{sect:spscalability}) and classification (\ref{sect:spml}).

\subsection{Scalable Single Pulse Identification}
\label{sect:spscalability}

The granularity of SPEs considered in this paper is much finer than that of DPGs. For example, the version of RAPID described in \cite{mnras} considers only the maximum SNR for each DM and could only identify one DPG in the data represented in Figure~\ref{fig:spplot1}. Our new distributed, single pulse version of RAPID identified 188 single pulses in the same data, including \textit{single pulse\#1} and \textit{single pulse\#2}. Consequently, examining the data at this granularity is several orders of magnitude more processor intensive. For DPG identification, RAPID needs to run once for each observation. D-RAPID for single pulse identification, on the other hand, must run once for \emph{every cluster of SPEs} in an observation. Since most observations contain several hundred to several thousand clusters, the modified algorithm must run several hundred to several thousand more times for each observation. Considering one dataset used in our experiments encompasses almost 300 million observations, the workload is too much for a single computer to finish in a reasonable amount of time, even with multithreaded programming. Fortunately, the problem of single pulse identification may easily be divided into many smaller subproblems which can then be solved simultaneously. The independence of the data combined with the uniform processing requirement makes searching for single pulses ideal for data parallelism on a distributed computing platform. Once a master node distributes the data to the data nodes of the distributed system, each data node performs the same sequence of operations on its local data, returning the results to the master for aggregation.

We designed scalability into our solution by parallelizing our application to run on a distributed YARN cluster with Apache Spark. YARN is a resource management and job scheduling technology that makes two key improvements over the traditional Hadoop design. First, YARN decouples the programming model from the resource management infrastructure to allow programmers more flexibility for diverse coding applications. Second, YARN delegates many scheduling functions (e.g., task fault-tolerance) to per-application components, effectively decentralizing the management of each job's control flow and thus improving scalability and efficiency \cite{Vavilapalli:2013:AHY:2523616.2523633}. Another advantage of a YARN architecture is its compatibility with Apache Spark, which is a fast and general distributed processing engine that has been shown to outperform the more traditional Hadoop MapReduce framework by ten times in iterative machine learning jobs \cite{Zaharia:2010:SCC:1863103.1863113}. Spark was designed for problems that need to reuse a working set of data across multiple parallel operations. To optimize such tasks, Spark introduced the resilient distributed dataset (RDD), which is a collection of objects partitioned across a set of data nodes that can be rebuilt if a partition is lost. Spark achieves significant speed improvements by providing built-in functions to transform RDDs \emph{in memory}, whereas MapReduce requires each intermediate step to be written to the HDFS. These qualities make a Spark on YARN cluster well-suited for the problem of efficiently identifying single pulses in massive data sets.

\subsubsection{D-RAPID}
\label{sect:DSPID}

To take full advantage of distributed computing with Spark, we designed our data flow using RDDs. We wrote the original driver code behind RAPID in Java. For the redesign, we switched to the hybrid functional and object-oriented programming language Scala, which is the development language of choice for Spark applications \cite{Odersky:2011:PSC:1983576}. Using Scala to develop our driver allowed for the seamless integration of our existing Java code and the subsequent single pulse modifications described in Section~\ref{sect:sprapid}. In this paper, we annotate our distributed implementation of RAPID for single pulses in Scala as D-RAPID.

\begin{figure}[ht]
	\centering
	\includegraphics[width=0.55\linewidth]{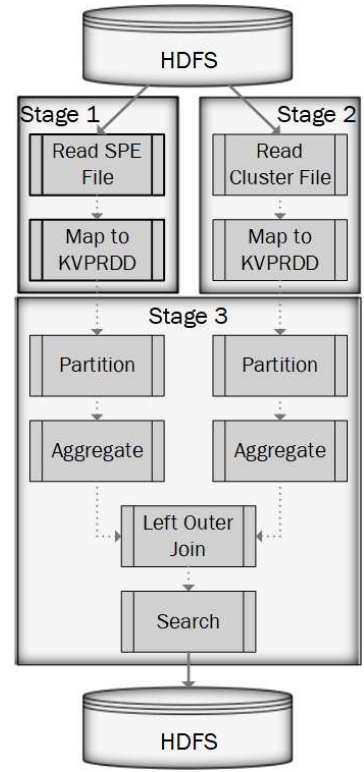}
	\caption{The stages of execution for D-RAPID to search a set of clusters for single pulses.}
	\label{fig:sparkdag}
\end{figure}

D-RAPID requires two input files to be loaded into the HDFS: a large data file containing all of the SPEs for an entire data set in csv format, and a smaller cluster file detailing each cluster to be searched (see Figure~\ref{fig:singlepulseflow}). The goal of D-RAPID is to search for single pulses \emph{only in the areas of the data file that coincide with the clusters listed in the cluster file}. This poses a challenge for a distributed environment because the cluster information and the data belonging to it may not exist on the same data node. In the HDFS, a single file may be split into many chunks and replications and stored on several different data nodes. To solve this problem, D-RAPID organizes and joins the cluster and data files into a single RDD in memory to guarantee that each executor has all of the information it needs locally. This process is accomplished in three stages, as shown in Figure~\ref{fig:sparkdag}. 

Stage 1 and Stage 2 represent loading and preparing the input files. We first stripped the files of their header information. Then, in the \textit{Map to KVPRDD} phase, we read each file from the HDFS and convert it into a Key-Value Pair RDD (KVPRDD), which can take advantage of the Spark's efficient built-in transformation functions. Since every instance in both the data and cluster files begins with the same descriptive information, i.e., the name of the data set, the mean Julian date (MJD) of the observation, its sky position, and the beam, we concatenated these descriptors to serve as the key for each instance. The value paired with each key is the remainder of the string containing the data for that instance, either SPE data or data describing a particular cluster. In Stage 3, we partition the data and cluster KVPRDDs and aggregate them to prepare for joining. D-RAPID uses Spark to optimize this process in two ways: uniform partitioning and key aggregation.

When joining two KVPRDDs, the key-value pairs must be shuffled and compared until they find their matches, which would result in an excessive amount of network traffic and overhead. Instead, we partition each KVPRDD in the exact same manner, so that the matching keys for each set are naturally colocated, eliminating unnecessary shuffling when performing the join operation\cite{highperfspark}. In the \textit{Partition} phase, D-RAPID uses a Spark HashPartitioner to hash the keys and shuffle partitions with the same keys to the same executors from both data sets.

A join performed between two RDDs with many duplicate keys can significantly inflate the size of the joined data set \cite{highperfspark}. An artifact of the csv format used for the data file is that there will be a very high number of repeated keys. In the \textit{Aggregate} phase, we aggregate the cluster and data KVPRDDs by key so there will be less pairs when the expensive join operation is performed. 

With the KVPRDDs optimized, we perform the join in the \textit{Left Outer Join} phase. Generally speaking, a left outer join on two data sets will return a value for every entry of the first data set, even if there are no matching entries in the second data set (in which case a null value is returned) \cite{Ramakrishnan:2002:DMS:560733}. The end result of the left outer join of our data sets is a new KVPRDD with one KVP for each cluster together with all of the SPE data from the data file necessary to process it. This ensures that when we apply D-RAPID to the joined pairs in the \textit{Search} phase, all necessary data will be local to the data node performing the operation. In the search phase, D-RAPID searches the SPE data from each cluster for single pulses. Finally, we write all identified single pulses back to the HDFS in separate files, which we later extract and from the HDFS and concatenate in preparation for the classification phase.

\subsubsection{Single pulse Identification with D-RAPID}
\label{sect:sprapid}

Identifying single pulses in the DM vs. time space instead of aggregated DPGs in the SNR vs. DM space required several modifications to our previous searching algorithm \cite{mnras}. We developed Algorithm~\ref{alg:drapid} to identify single pulses in the search phase of D-RAPID. The search algorithm uses recursion to divide the SPEs in a cluster into bins, performs a linear regression on the points in each bin to determine trends, and uses the trends to identify peaks in the data. For each bin, we consider the trend of the previous bin, $b_{n-1}$, which can be either decreasing, flat, or increasing. We next consider the slope of the current bin, $b_n$ and the state of a potential single pulse candidate, $SP$, to determine whether the SPEs in the current bin are ``climbing'' a single pulse, have reached the peak, or are ``descending'' a single pulse.

\begin{algorithm}[t]
	\SetAlgoNoLine
		\KwIn{$start$ the index of the first SPE in a bin}
		\myinput{$b_{n-1}$ $\gets$ regression slope of the previous bin, initialized to 0}
		\myinput{$SP$ a potential single pulse, initialized to NULL}
		\myinput{$M$ the slope threshold, a constant}
		\KwOut{data mining file listing all identified single pulses}
		$next$ $\gets$ $start$ + $binsize$\;
		\uIf{$next$ > total number of SPEs}{\Return}
		Calculate $Y_i=a+bX_i+e_i$, a linear regression using all points in the current bin\;		
		\uIf{$b_{n-1}< -M$}{
			\uIf{$-M < b_{n} < M$ {\normalfont and} $SP = NULL$ {\normalfont or} $SP$ has no peak}{$SP \gets NULL$ and begin a new $SP$\;}
			\uIf{$b_n > M$ {\normalfont and} $SP != NULL$ and has a peak}{add this $SP$\;}
			begin a new $SP$\;
		}
		\uElseIf{$-M < b_{n-1} < M$}{
			\uIf{$b_n < -M$}{
				\uIf{$SP != NULL$ but has no peak}{peak found for this $SP$\;}
				\uElseIf{$SP = NULL$}{$SP\gets NULL$  and begin a new $SP$\;}
			}
			\uIf{$-M < b_{n} < M$ {\normalfont and} $SP != NULL$ and has a peak}{write this $SP$ and begin a new $SP$\;}
				\uElse{$SP \gets NULL$\;}
			\uIf{$b_n > M$}{
				\uIf{$SP = NULL$}{begin a new $SP$\;}
				\uElseIf{$SP$ has a peak}{write this $SP$ and begin a new $SP$\;}
			}
		}
		\uElseIf{$b_{n-1} > M$}{
			\uIf{$b_n < -M$}{peak found for this $SP$}
			\uElseIf{$-M < b_{n-1} < M$ {\normalfont and} $SP = NULL$}{begin a new $SP$\;}
			\uElseIf{$b_n > M$ {\normalfont and} $SP = NULL$}{begin a new $SP$\;}
			}
		search($next$, $b_n$)\;
\caption{D-RAPID search}
\label{alg:drapid}
\end{algorithm}

The search algorithm has several parameters that also required modification. The \textit{bin size} parameter determines how many consecutive SPEs will be included in each regression calculation. For DPG identification in \cite{mnras}, we kept the bin size fixed at 25, a value that was chosen for its favorable experimental performance. However, a static bin size is not suitable when running on clusters, because they vary in size from several SPEs to thousands of SPEs. A static bin size of 25 will put all SPEs in small clusters into one bin, making it impossible for D-RAPID to identify a peak.

To accommodate this range of cluster sizes, in D-RAPID we assigned the bin size dynamically according to Equation~\ref{eq:binsize}, where $n$ is the number of SPEs in a cluster and $w$ is the \textit{weight}, a new parameter which governs how quickly the bin size grows as the cluster size increases.

\begin{equation}
\label{eq:binsize}
bin size =\left\{
  \begin{array}{@{}ll@{}}
    1, & \text{if}\ n < 12 \\
    \lfloor w\sqrt{n} \rfloor, & \text{otherwise}
  \end{array}\right.
\end{equation}

Using a bin size of one for small clusters simply ``connects the dots'' by considering the difference between two points. Varying the bin size by the square root of the cluster size ensures that the bin size increases quickly at first, and then levels out as cluster sizes become larger. The square root function exemplifies this behavior, but was found to increase too quickly for smaller clusters. To control the growth of the bin size, we introduced the weight parameter.

To tune the weight, $w$, and \textit{slope threshold}, $M$, parameters, we chose several single pulses that are difficult to identify from known pulsars and used them for parameter tuning. The slope threshold provides a minimum slope requirement to distinguish between a linear regression line that is flat, or is increasing (or decreasing). In the parameter tuning experiment, we allowed the weight to vary from 0.75 to 1.75 and the slope threshold from 0.05 to 0.5. The results showed that the combination of a weight of 0.75 and a slope threshold of 0.5 most efficiently identified problematic single pulses, and we used this parameter combination for the rest of the experiments.

\subsubsection{Feature Extraction}
\label{sect:spfeatures}

In \cite{mnras}, we described the extraction of a set of 16 features to be used by classifiers to distinguish pulsar single pulses from non-pulsar single pulses. Focusing on individual clusters of SPEs allowed the inclusion of several additional features, which are given in Table~\ref{tab:spfeatures}. The first four features are specifically defined for single pulses and are calculated from the data used to create DM vs. Time subplots, like the one in Figure~\ref{fig:spplot1}. The last two features could be extracted for either DPGs or single pulses.

\begin{table}[ht]
	\caption{Additional features extracted for each cluster and used by machine learning algorithms for classification. }
	\centering
	\footnotesize
	\begin{tabular}{ll}
	\toprule
	\textbf{Feature} & \textbf{Description} \\
	\midrule
		\textit{StartTime} & The arrival time of the first SPE in the cluster. \\
		\textit{StopTime} & The arrival time of the last SPE in the cluster. \\
		\multirow{2}{*}{\textit{ClusterRank}} & An SNR based ranking of the cluster compared to others \\
		 & in the same observation (as in \cite{di}). \\
		\multirow{2}{*}{\textit{PulseRank}} & The rank of a particular peak when compared to other \\
		 & peaks found in a cluster and ordered by SNRMax. \\
		\textit{DMSpacing} & The interval between two consecutive DM values. \\
		\multirow{2}{*}{\textit{SNRRatio}} & The ratio of the SNR of the first point in the peak to the \\
		& maximum SNR. \\
	\bottomrule
	\end{tabular}
	\label{tab:spfeatures}
\end{table}

We included the \textit{DMSpacing} because it affects how single pulses or DPGs appear in different DM ranges, as noted in \cite{di}. The \textit{DMSpacing} is the minimum difference between two consecutive DM values, and increases from 0.01 for low DM values to 2.00 for very high DM values in the DM vs. time space. Since D-RAPID calculates the slope of a linear regression through the points of a bin, differences in scaling on the DM-axis should also be taken into consideration when selecting a minimum slope threshold. In our parameter tuning trials we found that a slope threshold of 0.5 identified single pulses well regardless of the \textit{DMSpacing}. The \textit{SNRRatio}, a normalized ratio of the maximum SNR value to the SNR of the first point in a singlepulse, proved to be a discerning attribute when ranking attributes by their information gain.

\subsection{Single Pulse Classification with Machine Learning}
\label{sect:spml}
In this section, we first discuss our choices for dealing with imbalance in the training data. Then, we introduce a new data-driven, automatic multiclass classification scheme for single pulses. We next discuss how we identified the most relevant features to include in machine learning by using five different feature selection methods. Finally, we describe the performance measures we used to evaluate the effectiveness of our methods. By combining our novel multiclass approach with feature selection, we achieved significant improvements to the execution performance of machine learning algorithms while maintaining comparable classification performance.

\subsubsection{Imbalance Treatment}
\label{sect:spimbalance}

A data set is considered to be imbalanced when a large majority of its instances belong to only a few classes. In the data set we used for single pulse classification, only 0.05\% of the instances are from known pulsars. A supervised machine learning algorithm trained on imbalanced data will inevitably become biased toward the majority classes and more likely to miss the minority classes. While imbalance is still prevalent in our single pulse data set, examining SPEs rather than aggregated DPGs helped alleviate training bias by supplying more positive examples. Since a candidate plot can contain many single pulses from a single pulsar, we were able to identify ten times more positive training examples of single pulses from known pulsars than we could for DPGs \cite{mnras}. This led us to expect better performance from single pulse classifiers trained on benchmark data sets with no imbalance treatment. However, based on our favorable previous results, we also tested classifiers using Synthetic Minority Oversampling TEchnique  (SMOTE), which oversamples with small, random changes to the oversampled instances to avoid overfitting \cite{chawla2002smote}. To further avoid overfitting, we did not apply SMOTE to any of our testing sets.

\subsubsection{Multiclass Classification}
\label{sect:spmulticlass}

Traditionally, pulsar classification is binary, i.e., a candidate plot either contains a pulsar, or it does not. However, candidate plots containing pulsars can appear drastically different, both to the eye and to a machine learning algorithm. The differences in their appearances reflect their different physical properties, like distance from the observer and signal strength, as well as artifacts of data processing.

In \cite{mnras}, we performed multiclass classification by manually dividing the positive examples of DPGs in our benchmark into four distinct classes. This visually-based multiclass classification scheme was more successful at detecting rare events with distinctive features (RRATs) than its binary counterpart. However, it had a significant drawback: the positive instances must be manually sorted by a human with a trained eye. This limitation is magnified in the case of single pulse classification, which yields many more positive examples.

\begin{table}[ht]
	\caption{The features and thresholds used for Automatically Labeled Multiclass Classification.}
	\centering
	\footnotesize
	\begin{tabular}{lcl}
 	\toprule
		\textbf{Feature} & \textbf{Threshold} & \textbf{Label} \\ 
  	\midrule
 		\multirow{3}{*}{\textit{SNRPeakDM}} & $[0, 100)$ & near \\ 
		& $[100, 175)$ & mid \\
		& $[175, \infty)$ & far \\
		\multirow{2}{*}{\textit{AvgSNR}} & $[0, 8]$ & weak \\ 
		& $(8, \infty)$ & strong \\ 
	\bottomrule
	\end{tabular}
	\label{tab:classes}
\end{table}

To overcome this limitation and retain the benefit of learning from differences in groups of single pulses, we developed a new Automatically Labeled Multiclass (ALM) classification. Rather than using visual appearance, we categorized single pulses by two of their extracted features: \textit{SNRPeakDM}, which is the DM value for the brightest SPE in the single pulse, and \textit{AvgSNR}, which is the average brightness of all of the SPEs in the single pulse.  This was accomplished by discretizing these numeric features and using their combinations to define class labels. We wrote code to automatically label known single pulses based on the thresholds given in Table~\ref{tab:classes}. For comparison, we labeled each benchmark using five different class labeling schemes, as shown in Table~\ref{tab:labels}. Each scheme is named by the number of classes it contains. Note that the scheme \textit{4*} is identical to the classification scheme used in \cite{mnras}. Also, to help determine if the new technique could still effectively classify rare instances, scheme \textit{8} includes RRATs as a separate class.

\begin{table}[ht]
	\caption{The five different multiclass classification schemes tested.}
	\centering
	\footnotesize
	\begin{tabular}{cl}
	\toprule
		\textbf{Scheme} & \textbf{Classes} \\ 
  	\midrule
		{\textit{2}} & Binary: Non-pulsar, Pulsar \\
		{\textit{4*}} & Non-pulsar, Pulsar, Very Bright Pulsar, RRAT (as in \cite{mnras}) \\
		{\textit{4}} & Non-pulsar, Near, Mid, Far \\
 		\multirow{2}{*}{\textit{7}} & Non-pulsar, Near-Weak, Near-Strong, Mid-Weak, Mid-Strong, \\
 		 & Far-Weak, Far-Strong \\
 		\multirow{2}{*}{\textit{8}} & Non-pulsar, Near-Weak, Near-Strong, Mid-Weak, Mid-Strong, \\
 		& Far-Weak, Far-Strong, RRAT \\
	\bottomrule
	\end{tabular}
	\label{tab:labels}
\end{table}

The \textit{SNRPeakDM} was chosen as a distinguishing feature for several reasons. The DM of a SPE is the integrated column density of free electrons along the line of sight between the observer and the SPE. In other words, if a long, narrow tube extended from the observer to the origin of the SPE, the DM would be proportional to the number of free electrons inside the tube. Theoretically, given the Galactic distribution of free electrons, the DM could provide the distance to the SPE and the \textit{SNRPeakDM} the distance to the strongest SPE of a single pulse \cite{handbook}. As a theoretical measure of distance, the \textit{SNRPeakDM} makes sense as a possible categorical feature. Additionally, by using \textit{SNRPeakDM} as a distinguishing feature we allow machine learning algorithms to take advantage of differences in other features that depend on DM, such as the \textit{DMSpacing}. Finally, through examination of the distribution of known single pulses in the benchmark data, the \textit{SNRPeakDM} was found to adequately group the known single pulses.

A key advantage of single pulse over DPG classification is its sensitivity to weaker signals, as measured by the SNR of a SPE. The \textit{AvgSNR} represents the relative, average strength or `brightness' of a single pulse, making it a potentially promising distinguishing feature. Furthermore, examination of known single pulses in the benchmark data revealed that several other metrics appeared to change proportionally to the \textit{AvgSNR}, such as the number of pulses.

\subsubsection{Feature Selection}
\label{sect:spfs}

In \cite{mnras} and Section~\ref{sect:spfeatures}, we described the extraction of 22 characteristic features of a single pulse. In practice, not all features are relevant to classification. Irrelevant features may reduce classification performance and typically increase computational costs and memory usage of classifiers \cite{BLUM1997245}.

In machine learning, feature selection is the process of removing irrelevant features from a data set. Feature selection methods can be either \textit{filters} or \textit{wrappers}. Filters are feature selection methods independent of the machine learning algorithms. They rely on statistical or heuristic selection criteria, like correlation or entropy measures, to choose the best features \cite{John:1994:IFS:3091574.3091590}. On the other hand, wrappers use the results of machine learning algorithms to perform feature selection. They greedily search the feature space for different combinations of features and evaluate the effectiveness of subsets by the classification performance of a given algorithm \cite{kohavi1997wrappers}.

\begin{table}[ht]
	\caption{The five different feature selection algorithms used.}
	\centering
	\footnotesize
	\begin{tabular}{ll}
	\toprule
		\textbf{FS Algorithm} & \textbf{Type} \\ 
	\midrule
 		{\textit{InfoGain} (IG)} & Entropy Measure \\
	 	{\textit{GainRatio} (GR)} & Entropy Measure \\
	 	{\textit{SymmetricalUncertainty} (SU)} & Entropy Measure \\
	 	{\textit{Correlation} (Cor)} &  Linear Correlation \\
	 	{\textit{OneR} (1R)} & Machine Learning \\
	\bottomrule
	\end{tabular}
	\label{tab:fs}
\end{table}

We independently applied five filtering feature selection techniques listed in Table~\ref{tab:fs} to our data to explore what effect, if any, they had on performance. Each feature selection method provides a ranking for each feature which can be used to select only the top-ranked features.  The implementations of each feature selection method are available through \textit{Weka} \cite{weka}.

\subsubsection{Performance Measures}
\label{sect:spmetrics}
We evaluated classification performance by constructing \textit{confusion matrices} and then calculating the Recall, Precision, and F-Measure values for each classifier. A confusion matrix is a summary table of a classifier's performance on a given test set that categorizes each classification as one of the following: 
 
\begin{itemize}
	\item \textit{True Positive} (TP) -- a correctly classified pulsar single pulse.
    \item \textit{True Negative} (TN) -- a correctly classified non-pulsar single pulse.
	\item \textit{False Positive} (FP) -- an incorrectly classified non-pulsar single pulse.
	\item \textit{False Negative} (FN) -- an incorrectly classified pulsar single pulse.
\end{itemize}

Recall measures the ability of a classifier to correctly classify positive training instances, and is defined as: 
\begin{equation}
Recall = \frac{TP}{TP + FN}.
\end{equation}

Precision describes what fraction of the positive classifications are relevant, and is defined as:
\begin{equation}
Precision = \frac{TP}{TP + FP}.
\end{equation}
A low precision indicates that a classifier falsely labeled many non-pulsars as pulsars, resulting in a large number of instances requiring manual inspection.

The F-Measure is the harmonic mean of the Recall and Precision of a classifier, and is defined as:
\begin{equation}
F\mbox{-}Measure = 2*\frac{Precision*Recall}{Precision + Recall}.
\end{equation}
In effect, the F-Measure characterizes a classifier's ability to both not miss any pulsars and minimize the number of false positives produced \cite{Witten:2005:DMP:1205860}. While a high Recall is important for any classifier, the goal of automated classification is to minimize human involvement. A low F-Measure score with a high Recall would indicate that a classifier generates many false positives which will require manual inspection. For automated pulsar classification, both high Recall and F-Measure scores are desirable. 

Additionally, we consider the execution performance of each classifier by comparing their training times. Testing times are not reported in this work. While testing times are important for the application of classifiers, this work is focused on performance in an experimental environment given small subsets of the full survey data. We hope to evaluate testing times on a production environment in future work.
 
\section{Empirical Evaluation}
\label{sect:cs-palfa}
In this section, we present an application of our single pulse identification and classification methodology to the real-world single pulse search data from the GBT350Drift and PALFA sky surveys described in Section~\ref{sect:spdata}. 

\subsection{Identification Results}
\label{sect:spscalabilityresults}

To demonstrate and evaluate the performance of our distributed approach to single pulse identification, we implemented our own distributed experimental environment and used it to empirically evaluate the performance of our scaled-up D-RAPID algorithm.
 
The experimental environment consisted of sixteen computers donated by Fairmont State University. Eight computers came equipped with a 3.2 GHz quad-core Intel\textregistered Core\texttrademark i5-3470 CPU, a 250 GB hard drive, and 8 GB of RAM, and the remaining eight machines had Intel\textregistered Core\texttrademark2 Duo E8600 CPUs clocked at 3.33 GHz, 225 GB hard drives, and 4 GB of RAM. We installed 64 bit Ubuntu 14.04 on each machine and upgraded one of the i5 machines with an additional 8 GB RAM to serve as the master node. The other fifteen computers were configured as data nodes. Overall, the distributed system provided 60 virtual cores, 115.74 GB of available RAM, and 3.2 TB of resilient data storage. We configured the distributed system to utilize the Hadoop YARN architecture, and managed it with the open source Apache cluster management software \textit{Ambari} \cite{Wadkar2014}. 

We evaluated D-RAPID in our experimental environment on a 10.2 GB subset of the full PALFA SPE data set, which led to a 200 MB cluster file containing infromation for over 1.9 million clusters identified by the modified DBSCAN clustering algorithm presented in \cite{di}. In the distributed experimental environment, each executor was given two virtual processor cores and 2,560 MB of RAM so that the testing environment could support a maximum of 22 executors. Our custom partitioner assigned 32 partitions for each core, for a total of 896 partitions of the data. We recorded the total elapsed time to finish processing the test data in five separate trials, controlling the number of executors so that 1, 5, 10, 15, and 20 executors were allowed to operate in parallel. 

\begin{figure}[ht]
	\centering
	\includegraphics[width=0.6\linewidth]{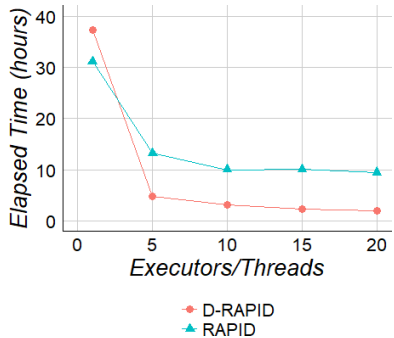}
	\caption{Performance evaluation of elapsed time to execute D-RAPID and a multithreaded version (RAPID) on a 10.2 GB test set with 5, 10, 15, and 20 executors or threads, respectively.}
	\label{fig:sparktimes}
\end{figure}

For comparison, we also tested a multithreaded version of RAPID for single pulse identification on the same subset of the PALFA data. The multithreaded version ran on a single processor machine with 16 GB of RAM and an Intel\textregistered Core\texttrademark i7-7800K CPU overclocked to 4.5 GHz. We processed the test data five times, allowing the program to use 1, 5, 10, 15, and 20 threads to accomplish the task. Figure~\ref{fig:sparktimes} provides the results.

\paragraph{RQ 1: Is D-RAPID scalable?}

The elapsed processing time for D-RAPID decreased as the number of executors increased, as Figure~\ref{fig:sparktimes} shows. The knee of the curve for D-RAPID occurs when given five executors, and the elapsed time decreases asymptotically as the number of executors grows beyond five. These results indicate that D-RAPID can utilize Spark on a YARN cluster to dramatically reduce processing times for single pulse identification. While this is a good result, the significantly smaller decrease in elapsed processing time when using more than five executors implies that other performance bottlenecks may exist.

One such bottleneck arises from the serial processing requirement for D-RAPID's search task, outlined in Algorithm~\ref{alg:drapid}, combined with the considerable size differences in various clusters. In the test data, individual clusters ranged in size from less than five SPEs to over 3,500 SPEs, with a median size of 19 SPEs. The sequential searching task took considerably longer for very large clusters. When examining the task distribution (for trials with more than one executor), some executors inevitably processed significantly less clusters due to these size differences.

The relatively small size of our test data set and the limitations of our experimental environment may have also impacted our performance results. We plan to further examine these potential bottlenecks in future work.

\paragraph{RQ 2: Does D-RAPID outperform its multithreaded implementation?}

D-RAPID processed the same amount of test data in 22\% to 37\% of the time it took its multithreaded counterpart in all trials, except when run with only one executor. A single executor, with only 2,560 MB of RAM, cannot fit the entire test data set into memory at once, which effectively eliminated the advantage of parallel processing with Spark, as portions of the RDDs must be frequently swapped out to disk to compensate for the lack of memory. This leads us to conclude that, as long as a YARN cluster has enough executors and memory to fit the entire data set into its distributed RAM, D-RAPID will consistently outperform the multithreaded RAPID implementation by up to a factor of five.

\subsection{Classification Results}
\label{sect:spbenchres}

For both PALFA and GBT350Drift, we created one benchmark data set for each of our five multiclass labeling schemes. We divided each of these ten benchmarks into six folds, one for feature selection and the other five for training and testing classifiers. We used each of our five feature selection techniques to rank the ten most relevant features from the first fold. Before classification, we removed features from the remaining five folds keeping only the top ten features as indicated by the given feature selection technique. We also kept a set of benchmarks with no feature selection as a baseline comparison. We then performed classification trials with the six machine learning algorithms, given in Table~\ref{tab:learners} with implementations available through \textit{Weka} \cite{weka}, using five fold cross validation on each benchmark for a total of 1,800 individual experiments. We repeated this process with another full set of benchmarks that we balanced with the SMOTE imbalance treatment, described in Section~\ref{sect:spimbalance}, for a total of 3,600 classification trials.

\begin{table}[ht]
	\caption{The name and type of each machine learning algorithm used for this work.}
	\centering
	\footnotesize
	\begin{tabular}{ll}
	\toprule
		\textbf{Learner} & \textbf{Type} \\
	\midrule
		MPN & Artificial Neural Network \\
		SMO & Support Vector Machine \\
		JRip & Rule \\
		J48 & Tree \\
		PART & Rule + Tree \\
		RandomForest (RF) & Ensemble Tree \\
	\bottomrule
	\end{tabular}
	\label{tab:learners}
\end{table}

\subsubsection{Multiclass Results}
\label{sect:results-mc}

The research questions in this section explore the effects of different Automatically Labeled Multiclass (ALM) schemes on single pulse classification performance and execution performance. Of the 3,600 experimental trials, only results of the 600 trials without feature selection are reported in this section.

Multiclass labeling scheme \textit{4*}, which is the scheme successfully used to label DPGs described in \cite{mnras}, exhibited the poorest performance overall. Due to significant visual differences between single pulses and DPGs, it is not surprising that a visually-based multiclass labeling scheme for DPG classification is ineffective for single pulse classification. Scheme \textit{4*} performs poorly throughout all experiments, and its results are omitted. 

\begin{figure*}[ht]
	\centering
	\begin{tabular}{c c }
		\includegraphics[width=0.45\linewidth]{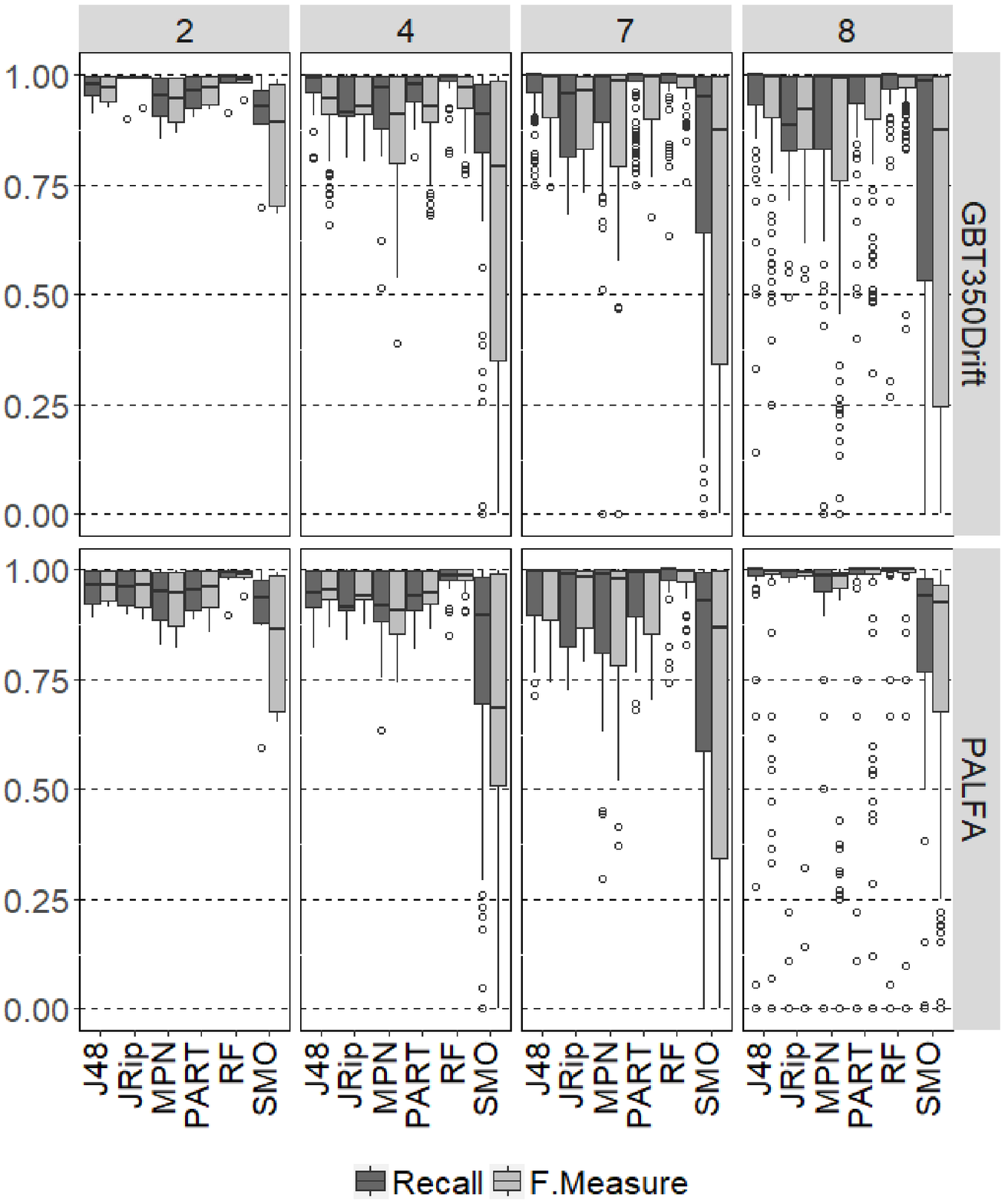} &
		\includegraphics[width=0.45\linewidth]{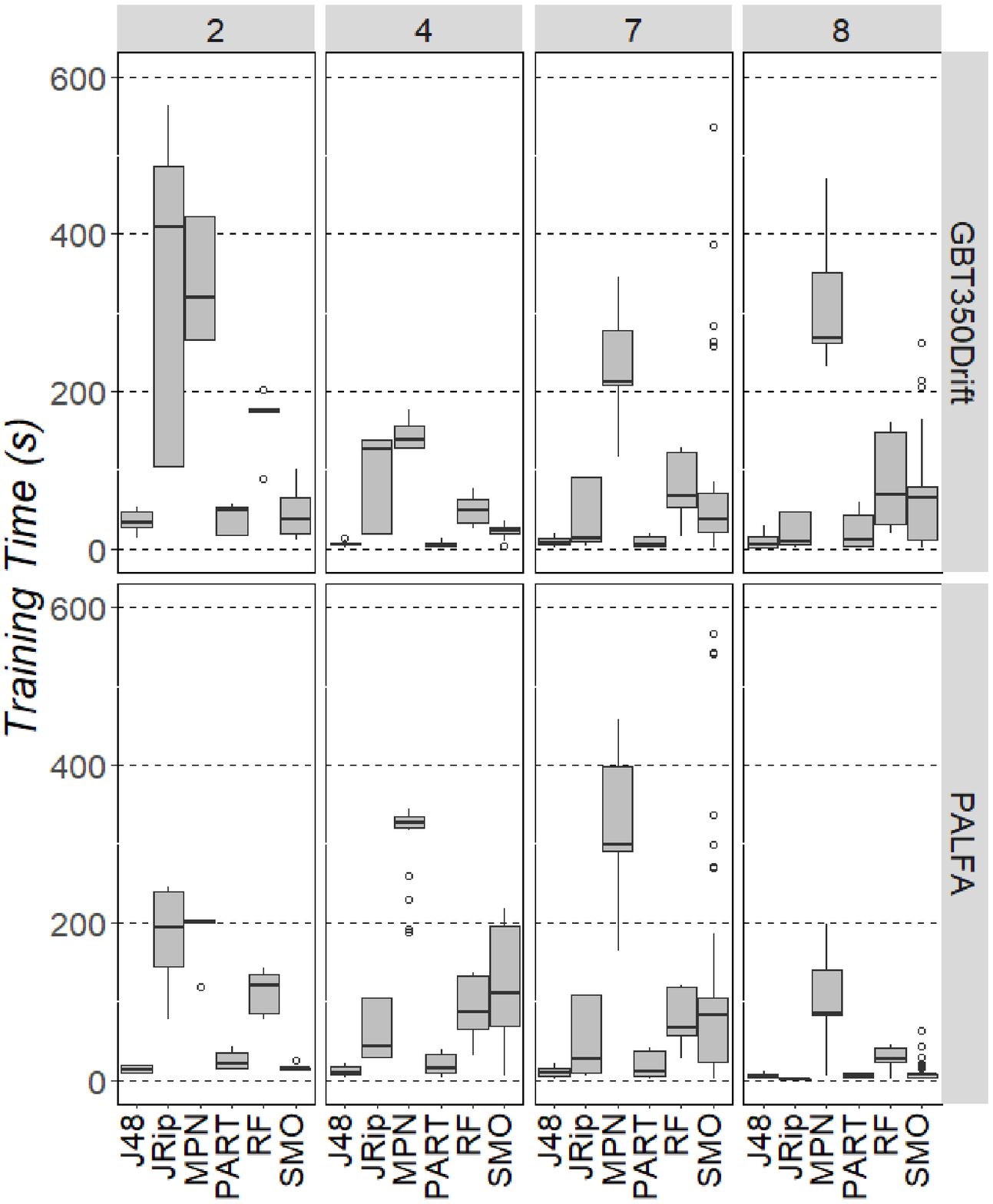} \\
		\textbf{(a)} & \textbf{(b)} \\
	\end{tabular}
\caption{Boxplots of the Recall, F-Measure scores (a) and training times (b) for all classifiers subdivided by ALM scheme (columns) and data set (rows).}
\label{fig:rq3}
\end{figure*}

\paragraph{RQ 3: Does ALM improve classification performance?}
\label{sect:results-mc-perf}

Figure~\ref{fig:rq3}(a) shows boxplots of the Recall and F-Measure scores for all classifiers subdivided by ALM scheme (columns) and data set (rows). The plots show that performance results for ALM scheme \textit{4} were comparable to those of binary classification for most algorithms. Binary RF classifiers resulted in the highest median Recall and F-Measure scores with the smallest inter-quartile ranges (IQRs), except for the PALFA RF classifier using ALM scheme \textit{8}, which performed the best overall.

\paragraph{RQ 4: Does ALM improve classification performance for rare events?}
\label{sect:results-mc-rare}

To explore this research question, we created a list of all positive instances of single pulses and which classifiers were able to correctly classify each of them. We used the list to determine which single pulses were mis-classified by the most classifiers. The twenty most mis-classified single pulses were missed by 90 - 99\% of all classifiers. We found that ALM classifiers were more than twice as likely to correctly classify these problematic instances than binary classifiers. Furthermore, when expanding the analysis to include single pulses missed by 75 - 99\% of classifiers, ALM classifiers were over three times more likely than binary classifers to make correct classifications. This analysis also showed that RF classifiers were far better at classifying problematic single pulses, as they accounted for more correct classifications than all other classifiers with different algorithms combined.

\paragraph{RQ 5: Does ALM improve execution performance?}
\label{sect:results-mc-time}

Figure~\ref{fig:rq3}(b) shows boxplots of training training times for all classifiers subdivided by ALM scheme (columns) and data set (rows). The boxplots show a noticeable reduction in median training times for J48, JRip, MPN, PART, and RF. Note that training times for SMO, which had the worst classification performance, had a high number of very large outliers as the number of classes increases, and a consistent increase in median training times. While RF classifiers exhibited the best overall classification performance, the simpler learners J48 and PART had the fastest training times. However, the long training times for RF classifiers were consistently reduced by the application of ALM. Overall, ALM scheme \textit{8}, which was automatically labeled by DM ranges and SNR strength, exhibited the fastest training times for RF (on average 56\% faster than binary RF classifiers). These results indicate that ALM improves the execution performance of classifiers while maintaining classification performance comparable to their binary counterparts.

Overall, ALM RF classifiers, on average, exhibited both Recall and F-Measure scores within 2\% of their binary counterparts. Average total training times, however, were 47\% less than binary RF classifiers. Due to these significant improvements in execution performance, combined with ALM RF's improved classification of difficult instances, ALM RF classifers appear to be the best choice for single pulse classification.

\subsubsection{Feature Selection Results}

The research questions in this section explore the effects, if any, of feature selection on single pulse classification performance and execution performance. This analysis includes results from all 3,600 classification experiments.

\begin{figure*}[ht]
	\centering
	\begin{tabular}{c c}
		\includegraphics[width=0.45\linewidth]{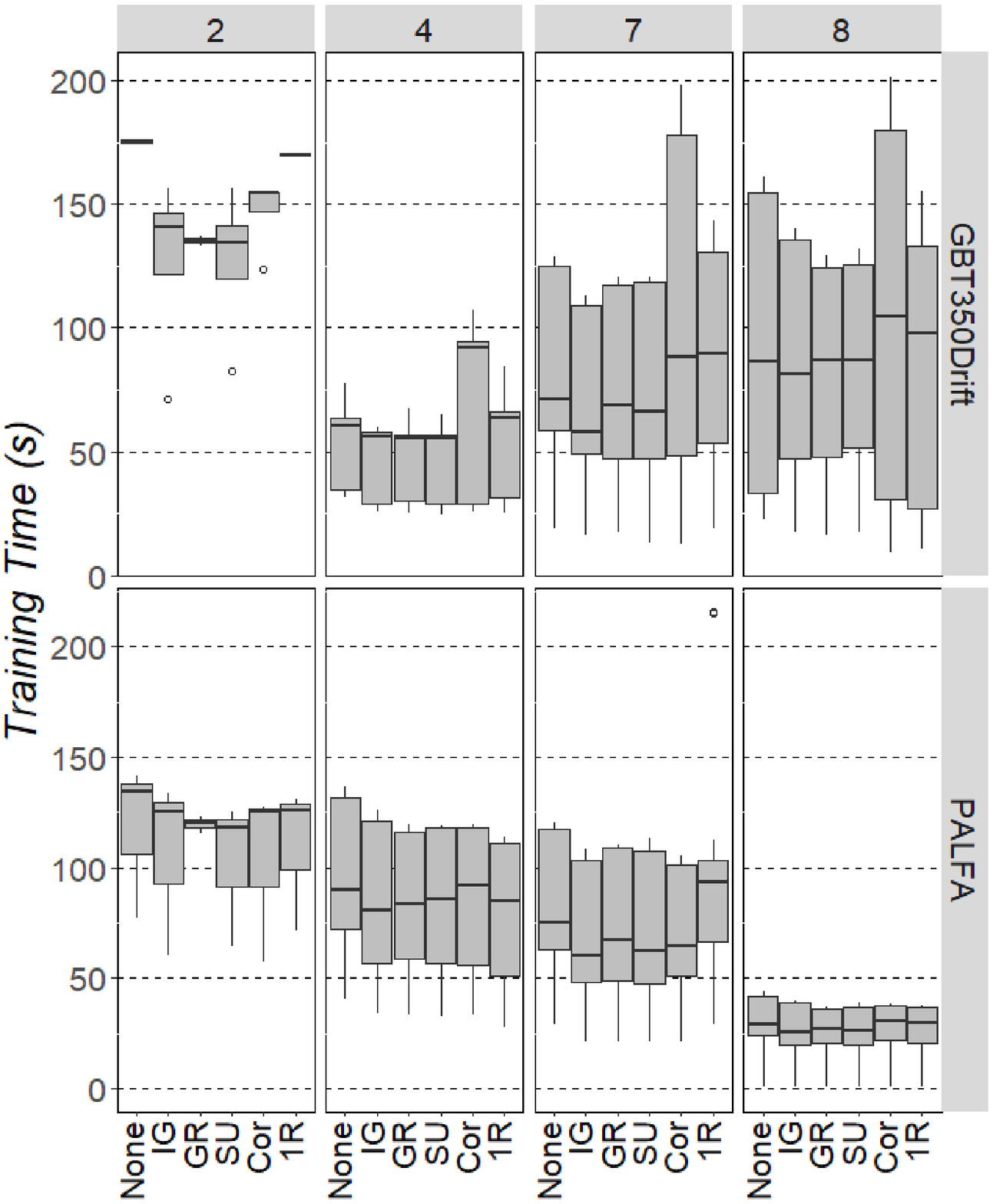} &
		\includegraphics[width=0.45\linewidth]{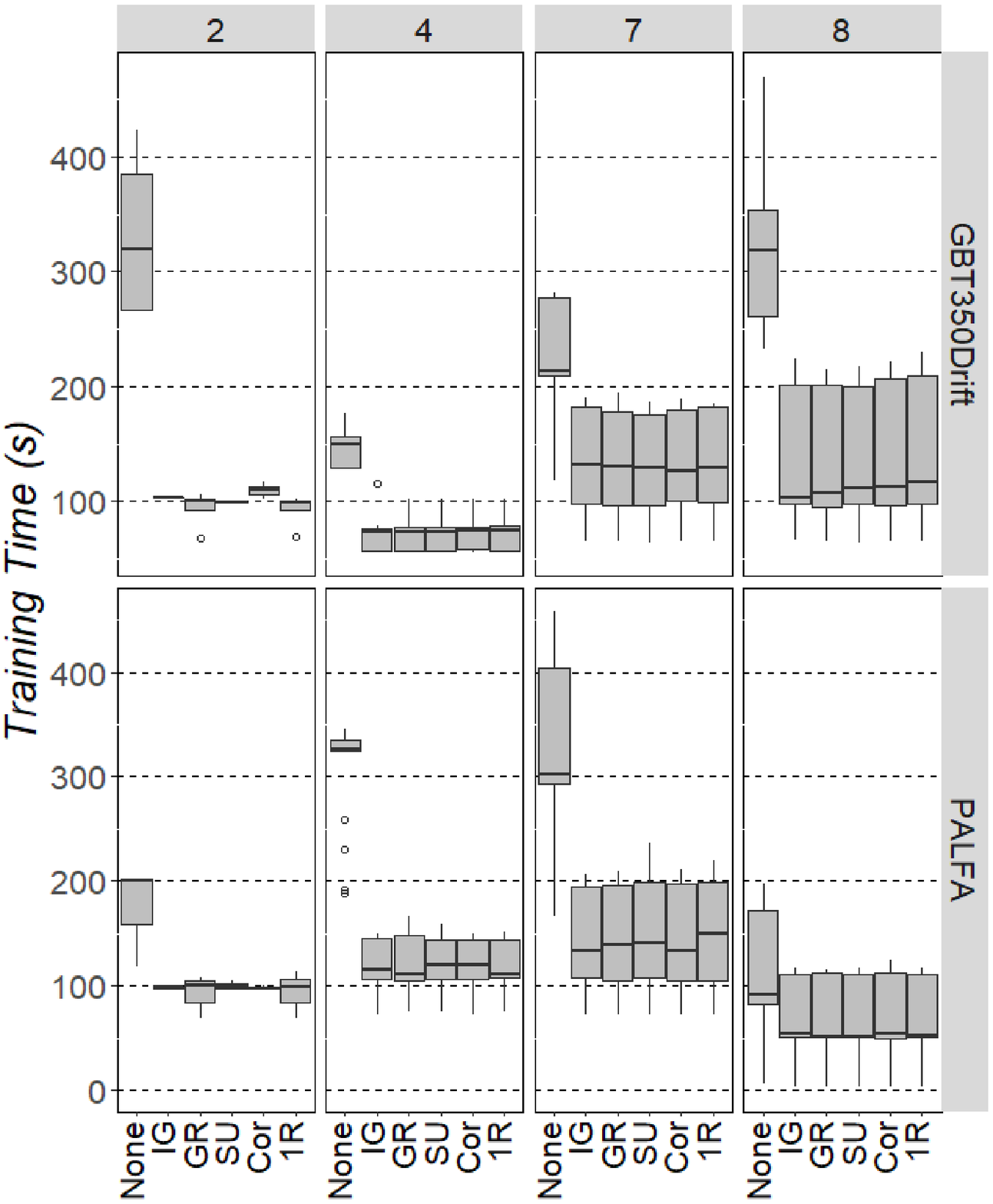} \\
		\textbf{(a)} & \textbf{(b)} \\
		\textbf{RF} & \textbf{MPN} \\
	\end{tabular}
\caption{Boxplots showing training times for RF (left) and MPN (right) classifiers. The x-axis shows the feature selection method and the subdivisions show the ALM scheme for the GBT350Drift (top row) and PALFA (bottom row) data sets. Note that the scales are different.}
\label{fig:rq6-time1}
\end{figure*}

\paragraph{RQ 6: Does feature selection improve single pulse classification performance?}
\label{sect:results-fs-perf}

The classification results (omitted for space considerations) showed no significant benefit or detriment of feature selection on classification performance. For the RF classifiers, which exhibited the best overall classification performance, feature selection with IG, GR, and SU appeared to have no impact on Recall or F-Measure scores, while scores had larger inter-quartile ranges (IQRs) and lower medians for Cor and 1R. Similarly, classification performance for MPN classifiers was not affected by IG and GR.

\paragraph{RQ 7: Does feature selection improve the execution performance of single pulse classification?}
\label{sect:results-fs-time}

Figures~\ref{fig:rq6-time1}(a) and \ref{fig:rq6-time1}(b) show boxplots of training times from classification trials organized into columns by ALM scheme and rows by data set, for RF and MPN classifiers. (Results for other classifiers omitted for space considerations.) The $x$-axis of each boxplot shows the five different feature selection methods described in Section~\ref{sect:spfs}, as well as the initial classifier with no feature selection, annotated with "\textit{None}" in Figure~\ref{fig:rq6-time1}.

Certain feature selection techniques consistently improved execution performance with respect to training times. For the PALFA data set, ALM scheme \textit{8} exhibited the most significant decrease with the least variability, while ALM scheme \textit{4} performed best for the GBT data set. For multiclass \textit{RF} classifiers, the \textit{InfoGain} feature selection technique resulted in consistently faster training times. Since Recall and F-Measure scores for multiclass \textit{RF} classifiers using \textit{InfoGain} were comparable to those with no feature selection, we conclude that \textit{InfoGain}, an entropy-based feature selection technique, decreased the execution time of single pulse \textit{RF} classifiers.

Additionally, for all \textit{MPN} classifiers, all feature selection techniques resulted in reduced training times, in some cases (such as PALFA scheme \textit{7}) by up to 200 seconds. On average, training times for IG binary MPN classifiers were 64\% lower than their counterparts without feature selection. The significantly longer training times of \textit{MPN} classifiers were their major drawback (note the difference in scale between Figures~\ref{fig:rq6-time1}(a) and \ref{fig:rq6-time1}(b)). Figure~\ref{fig:rq6-time1}(b) shows that feature selection could be an effective means of mitigating the poor execution performance of neural network classifiers.

\section{Conclusions}
\label{sect:spconc}

Searching for single pulses in radio astronomy is a processor-intensive task  that consists of identification and classification phases performed on very large data sets. In this paper, we presented a software solution to improve the execution time of single pulse pulsar identification and classification.

We improved the execution performance of the identification phase by developing Distributed Recursive Algorithm for Peak IDentification (D-RAPID), a Scala implementation of peak identification that achieves scalability by parallelizing the data on a Hadoop YARN distributed system for in-memory task processing with Apache Spark. We partitioned separate but linked data in memory to be colocated on data nodes. Experimental results show that D-RAPID executes up to five times faster than a variant which achieves parallelism through multithreaded execution.

To improve execution performance during the classification phase, we introduced a new Automatically Labeled Multiclass (ALM) classification technique which labels instances for supervised machine learning based on thresholds of key distinguishing features. We also used feature selection to reduce the dimensionality of our data and achieve execution performance improvements.

We tested the performance of ALM and feature selection by performing classification trials using all combinations of six supervised machine learning algorithms, five ALM schemes, and six feature selection methods. We conducted the trials on two different benchmark datasets, enhancing the generalization of our conclusions. The results showed that while binary classification performances were very good, several multiclass schemes produced comparable classification results (Recall and F-Measure scores within an average of 2\%) but with significant execution performance improvements. Overall, the ensemble tree RandomForest (RF) classifiers using ALM and the entropy-based InfoGain (IG) feature selection method achieved the best classification performance with average Recall and F-Measure scores of 0.96 and 0.95, respectively. The ALM classifiers were also much better at correctly classifying instances that were often mis-classified by binary classifiers. Further, the total execution performance of the classification improved on average by 54\%, (47\% from ALM and 7\% from IG) when compared to RF classifiers without ALM and feature selection.

The scalability and execution performance of pulsar identification and classification solutions must be considered to address the volume of data collected by the advanced radio telescopes of today and emerging technologies of the future. In future work, we plan to leverage distributed systems and parallel machine learning to further improve the execution performance of pulsar classification.

\begin{acks}
This work is partially supported by National Science Foundation Award No. OIA-1458952. We thank Maura McLaughlin for her expertise and assistance with this research and the PALFA collaboration for making these data available and for helpful discussions. We also thank the reviewers of this paper for their helpful comments.
\end{acks}

\bibliographystyle{ACM-Reference-Format}
\bibliography{TD_PhD_Dissertation}

\end{document}